\documentclass[sigconf]{acmart}

\usepackage{float}
\usepackage{multirow}
\usepackage{subcaption}
\usepackage{listings}
\usepackage{enumitem}

\AtBeginDocument{%
  }

\setcopyright{acmlicensed}
\copyrightyear{2025}
\acmYear{2025}
\acmDOI{XXXXXXX.XXXXXXX}

\acmConference[ICPE '25]{the 16th ACM/SPEC International Conference on Performance Engineering}{May 05--09,
  2025}{Toronto, Canada}
  
\acmISBN{978-1-4503-XXXX-X/18/06}
\acmSubmissionID{\#46}

\begin{document}

\title{Multi-Strided Access Patterns to Boost Hardware Prefetching}

\author{Miguel O. Blom}
\affiliation{%
  \institution{LIACS, Leiden University}
  \city{Leiden}
  \country{The Netherlands}}
\email{m.o.blom@liacs.leidenuniv.nl}

\author{Kristian F. D. Rietveld}
\affiliation{%
  \institution{LIACS, Leiden University}
  \city{Leiden}
  \country{The Netherlands}}
\email{k.f.d.rietveld@liacs.leidenuniv.nl}

\author{Rob V. van Nieuwpoort}
\affiliation{%
  \institution{LIACS, Leiden University}
  \city{Leiden}
  \country{The Netherlands}}
\email{r.v.van.nieuwpoort@liacs.leidenuniv.nl}

\renewcommand{\shortauthors}{Blom et al.}

\begin{abstract}
Important memory-bound kernels, such as linear algebra, convolutions, and stencils, rely on SIMD instructions as well as optimizations targeting improved vectorized data traversal and data re-use to attain satisfactory performance. 
On contemporary CPU architectures, the hardware prefetcher is of key importance for efficient utilization of the memory hierarchy. 
In this paper, we demonstrate that transforming a memory access pattern consisting of a single stride to one that concurrently accesses multiple strides, can boost the utilization of the hardware prefetcher, and in turn improves the performance of memory-bound kernels significantly.
Using a set of micro-benchmarks, we establish that accessing memory in a multi-strided manner enables more cache lines to be concurrently brought into the cache, resulting in improved cache hit ratios and higher effective memory bandwidth without the introduction of costly software prefetch instructions.
Subsequently, we show that multi-strided variants of a collection of six memory-bound dense compute kernels outperform state-of-the-art counterparts on three different micro-architectures.
More specifically, for kernels among which Matrix Vector Multiplication, Convolution Stencil and kernels from PolyBench, we achieve significant speedups of up to 12.55x over Polly, 2.99x over MKL, 1.98x over OpenBLAS, 1.08x over Halide and 1.87x over OpenCV.
The code transformation to take advantage of multi-strided memory access is a natural extension of the loop unroll and loop interchange techniques, allowing this method to be incorporated into compiler pipelines in the future.
\end{abstract}



\maketitle

\section{Introduction}
\label{sec:introduction}


Various fundamental compute kernels, among which dense linear algebra and stencils, exhibit a low arithmetic intensity and are therefore memory-bound on many modern micro-architectures.
The predominant optimization objective of such kernels is towards the efficient utilization of the available memory bandwidth (and thus the entire memory hierarchy).
This is commonly addressed through the use of SIMD memory operations, optimization of access patterns for data re-use and methods that improve vectorized traversal of user-defined datastructures.
In modern micro-architectures, hardware (cache) prefetch engines play a pivotal role in the memory subsystem.
By detecting regular access patterns -- strides -- generated by a program, hardware prefetch engines aim to bring data that will potentially be accessed in the near future to the CPU cache ahead of time.
This results in fewer cache misses and 
lower overall memory access latency. 
Contemporary architectures utilize \emph{multiple} of such prefetch engines, which can each detect patterns and prefetch the corresponding data~\cite{Falsafi2014}.

Many techniques have been described to optimize memory-bound kernels by targeting data re-use.
However, for a large class of kernels that do not re-use cached data, these techniques are ineffective and optimizations must be found for bringing the data to the cache. 
This can be achieved using software prefetching~\cite{Mowry1992,Ainsworth2016}, but this remains hard to apply due to difficulty in determining a suitable offset and locality type at compile-time. 
Moreover, software prefetching introduces additional instructions to compute kernels.
Another approach is to change the way datastructures are allocated or traversed to improve hardware prefetcher performance~\cite{Marin2013}.
However, to the best of our knowledge, increasing the number of strides generated by a program to prime multiple hardware prefetch positions has never been considered and analyzed in depth.

In this paper, we explore the optimization of memory-bound compute kernels with regular access patterns to take advantage of \emph{multiple} hardware prefetchers within a single CPU core.
We achieve this by transforming the computational loop, converting a memory access pattern along a single stride to one that concurrently accesses multiple strides.
We refer to this as \textbf{\emph{multi-striding}}.
Through a rigorous performance analysis 
we demonstrate that multi-striding boosts the utilization of the hardware prefetcher, in turn improving the performance of memory-bound kernels significantly.

Subsequently, we show for a set of six memory-bound compute kernels that multi-strided configurations can be found that outperform single-strided baselines on three different x86-64 micro-architectures.
The surveyed kernels comprise Dense Matrix Vector Multiplication, 3x3 Convolution, Stencil Computation, and four other kernels from PolyBench~\cite{polybench},
Finally, we present a comparison of our multi-strided compute kernels to state-of-the-art reference implementations.
The multi-strided kernels achieve significant speedups: up to 2.99x over Intel MKL, 1.98x over OpenBLAS, 1.08x over Halide~\cite{Halide}, and 1.87x over OpenCV.
Note that these speedups are achieved without low-level hand optimization, but solely through the transformation of the memory access pattern. 
Our transformation builds upon common compiler transformations including loop unrolling, loop interchange, and simple loop vectorization. 
As such, our methodology is suitable to be incorporated into compiler pipelines in the near future.
We will release our generated multi-strided compute kernels as well as the scripts used to generate the different configurations to the community as an open source artifact before publication of the paper.

The remainder of this paper is organized as follows.
Section~\ref{sec:related_work} 
discusses 
related work.
Section~\ref{sec:approach} provides an in-depth explanation of multi-striding, whereas Section~\ref{sec:analysis} presents an elaborate analysis of this method regarding prefetcher utilization.
We describe how compute kernels can be transformed to take advantage of hardware prefetchers in Section~\ref{sec:transformation}.
Section~\ref{sec:experiments} explores the effectiveness of multi-striding on isolated components of compute kernels, and secondly presents a comparison 
to state-of-the-art reference kernels.
Section~\ref{sec:conclusion} concludes this paper. 

\section{Related Work}
\label{sec:related_work}

Cache prefetching has been extensively studied in the literature from the perspective of hardware as well as software.
Within hardware, many types of prefetchers have been implemented~\cite{Falsafi2014,Panda2024}, such as next-line prefetchers based on constant distances between accesses~\cite{Smith1978},
constant stride prefetchers, or prefetchers based on patterns of multiple distances using perceptrons~\cite{Bhatia2019}.
More recent techniques allow the offset at which a prefetcher operates to be dynamically tuned to specific executions~\cite{Michaud2016}, dynamically prioritize useful predictors specific to the execution~\cite{Panda2016}, and merge similar memory access patterns to enable faster prefetch techniques to be employed~\cite{Jiang2022}.
Traditional hardware prefetching relies on regular access patterns. Support for irregular access patterns, as seen in graph processing, requires sophisticated techniques such as a global access history~\cite{Nesbit2004}. Moreover, hardware prefetch address predictors for specific problem domains have recently been investigated, such as for vectorized codes~\cite{MartnezPalau2024} or sparse matrix multiplications~\cite{Luo2024}.


Software prefetch hints provide a degree of explicit control over \emph{what} data is intended to be prefetched to \emph{where} in the cache hierarchy. 
These instructions can be added manually 
or can be injected by the compiler~\cite{Mowry1991,Mowry1992}.
As it is difficult to determine the most suitable offset and locality type used for prefetching, many automated techniques have been developed~\cite{Guttman2015}, including profile-guided methods~\cite{Jamilan2022}.
Domain knowledge can be used in issuing software prefetches, e.g. about data partitions~\cite{Ainsworth2016}.
Marin et al. diagnosed application prefetching performance by determining the number of streams in a program by means of simulation~\cite{Marin2013}. Based on this analysis, they modified data structure allocations to improve prefetcher friendliness in two cases but did not change the access pattern to enhance prefetcher utilization, unlike our work.

Hybrid techniques are thoroughly explored~\cite{Wang2003}, as hardware prefetches incur less overhead than software prefetches but offer less freedom in steering prefetches towards useful addresses.
Good understanding of prefetchers by the compiler and programmer is required for combining hardware with software prefetching to be effective~\cite{Vanderwiel2000,Lee2012,Kuhn2024}.
Here, machine learning has also been used to improve prefetcher effectiveness by deciding on disabling certain hardware prefetchers given a program~\cite{Rahman2015, Hiebel2019}.
Knowledge extracted from a kernel and its datastructures can also be used to configure specialized hardware accelerators to guide hardware prefetchers~\cite{Kumar2014}.
Baer and Chen proposed hardware-based support through reference prediction and instruction lookahead~\cite{Baer1991}, which they later turned into a hybrid technique combined with software prefetching~\cite{Chen1994}. 
More recently, an approach was described to allow the hardware prefetcher to identify pointers in advance by annotating load instructions with type hints~\cite{Dong2024}, and to embed support at the OS-level to bring more transparency about the cache to the user runtime to improve prefetcher predictions~\cite{Garg2024}.

Besides prefetching, loop optimizations are a well-known technique used by compilers to improve cache performance of compute kernels.
This comprises \emph{loop splitting}, to isolate specific computations~\cite{Fuchs2014}; \emph{loop interchange}, to reorder the loop's nesting order~\cite{Vanderwijngaart1998}; \emph{loop skewing}, to offset loops for aligned accesses that reduce memory pressure~\cite{Zhao2005}; and \emph{loop blocking} to enhance data reuse~\cite{Mehta2016,Sioutas2018}.
However, all of these techniques focus mainly on cache re-use, whereas in this paper we investigate improving prefetch performance for kernels \emph{that do not exhibit re-use}.
As we will describe in the next section, loop unrolling plays an important role in our methodology.
Capabilities of loop unrolling have widely been explored~\cite{Davidson1995,Koseki1997,Huang1999,Sarkar2000,Booshehri2013}, but focus mainly on cache reuse, where only few explorations even acknowledge the hardware prefetcher~\cite{Wolf1996}.

Research into optimization of specific memory-bound kernels include optimizing 2D convolution stencils using the AVX2 extension~\cite{Amiri2017}, or stencil computations in general~\cite{Dursun2009} and matrix-matrix multiplications~\cite{Hemeida2020} using loop optimizations and software prefetching.
None of these approaches considered changing the access pattern to better prime the hardware prefetcher.

\section{Approach}
\label{sec:approach}
In this section we provide a description of \emph{multi-striding} and how this 
leads to improvements in hardware prefetcher performance.


\textbf{Multi-striding} is a generic approach where a compute kernel is transformed to access multiple contiguous sequences of memory addresses, strides, to make more effective use of the available memory bandwidth when processing typically dense data structures.
The rationale is that we prime the available prefetchers to predict at \emph{multiple} positions, concurrently, when using multiple strides.
These positions can be distanced at the original rows of the datastructure, and, if not present, we can apply loop blocking to cause these multiple positions to appear.
Typically, loop unrolling is performed on the \emph{contiguous data axis}, the axis where successive values are stored adjacently.
However, by means of loop unrolling over any other axis, multiple sequences in the contiguous data axis will be traversed simultaneously within the loop body, transforming the execution to take advantage of multi-striding.
This implies that wherever loop unrolling is feasible, multi-striding can likely be applied -- either manually or by a compiler backend.



\begin{figure}
    \centering
    \includegraphics[width=0.95\linewidth]{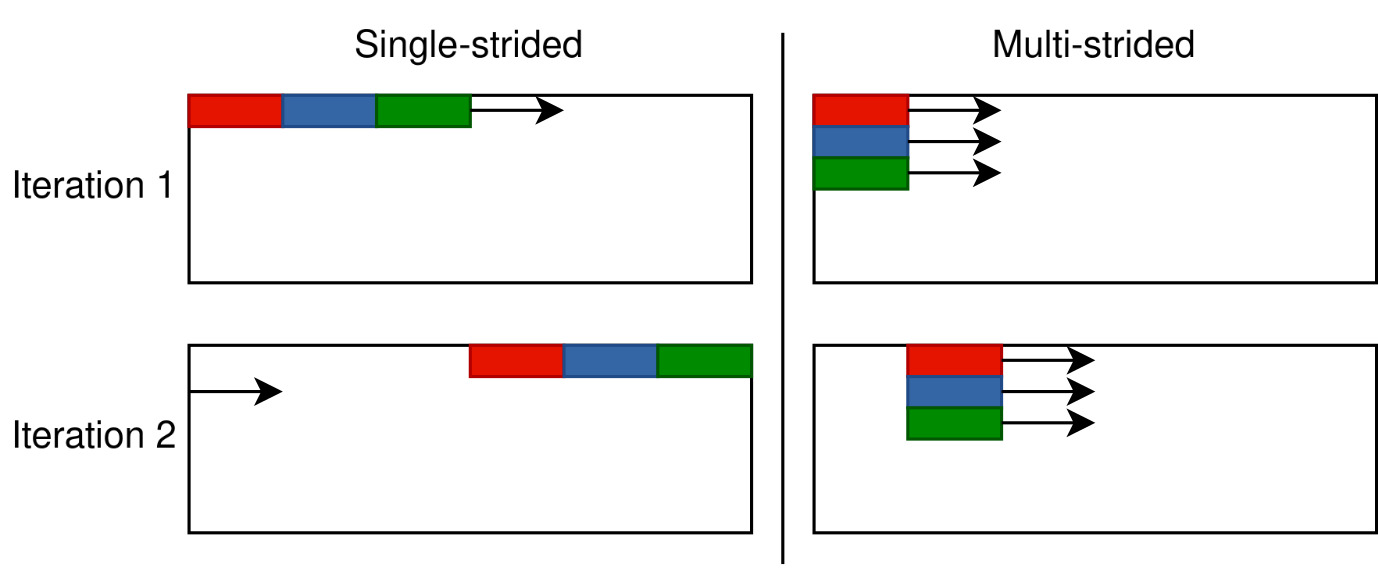}
    \caption{Illustration of the difference between a single-strided (left) and a multi-strided (right) access pattern, for two iterations, both using three unrolls.}
    \label{fig:striding}
\end{figure}

Figure~\ref{fig:striding} provides a visualization of how multi-striding differs from "normal" data access (i.e., single-striding).
The colored rectangles represent data accesses.
As can be seen, multi-striding involves loop unrolling across any other axis than the contiguous data axis, resulting in multiple strides to be iterated over \emph{simultaneously} rather than a single coalesced stride.
We define unrolling over the contiguous data axis, as seen in the single-strided configuration, as \emph{portion unrolling}, which leads to larger portions of each stride to be processed.
Unrolling over any other axis, as seen in the multi-strided configuration, is called \emph{stride unrolling}, which causes more strides to appear.
We can find an even distribution of $n$ loop unrolls over $d$ strides, as long as $d$ is a divisor of $n$, resulting in portions of length $n/d$.
We hypothesize that for every arrow in this diagram, a hardware prefetcher looks ahead at a fixed or variable length.
Thus, the overlap of an arrow in one iteration with an accessed value (rectangle) in the next iteration signifies that this value has already been prefetched and will be accessed with a lower latency, boosting the memory throughput. In this simple example, single-strided access prefetches only one element (one arrow), while the multi-strided version prefetches three elements (the three arrows in iteration 1 that overlap with the data used in iteration 2).

The performance improvement of multi-striding is to a large extent influenced by where data can be stored in the cache, that is the cache organization. The prevalent cache organization today is the set-associative cache. A $k$-way set-associative cache refers to a cache that is organized in sets where each set contains $k$ cache lines. In this placement policy, a cache block is assigned to a specific set 
based on its memory address.
Blocks spaced equally at a specific power of two are assigned to the same cache set. 
This means that accessing these blocks within a short time span increases the likelihood of too many strides competing for the same cache set.
This leads to conflict misses due to the eviction of cache lines, which can also affect prefetched blocks.
This is similar to the eviction of live cache blocks~\cite{Tambat2002,Seshadri2012,Bender2023}
with the addition that live \emph{prefetched} blocks are also evicted, compromising the effectiveness of the prefetcher. 

In reality, with the higher throughput vector instructions offer, data accesses are often vectorized.
So, each of the colored rectangles in Figure~\ref{fig:striding} actually represents a \emph{vector} of data elements. Within this work, we investigate several types of memory reads and writes using the AVX2 instruction set extension for single-precision floating point values. More in particular, we will consider:
\begin{itemize}[wide,noitemsep,labelwidth=!,labelindent=0pt]
\item Aligned reads and writes (\texttt{vmovaps}); these are regular memory operations and require a memory address aligned on vector size.
\item Unaligned reads and writes (\texttt{vmovups}); similar memory operations but allow unaligned memory addresses at the cost of the micro-architecture having to perform multiple memory accesses.
\item Non-temporal reads or streamed loads (\texttt{vmovntdqa}); this is an aligned memory read with the non-temporal hint, resulting in the backing cache line \emph{not} to be brought into the cache.
\item Non-temporal or streamed writes (\texttt{vmovntdq}); this is a write with the non-temporal hint, resulting in a write to memory without bringing the corresponding 
line into cache (no-write-allocate write).
\end{itemize}
\noindent Through the use of these different instructions, one can exert some control over what data is cached. For example, non-temporal store instructions avoid polluting the cache with non-reused data.
This is particularly beneficial for streaming large amounts of data that will not be reused. The write buffer temporarily holds data waiting to be written from the CPU to memory to allow for asynchronous (non-blocking) memory writes. Too many outstanding writes will overwhelm the write-buffer, as shown by Fritts~\cite{Fritts2000}, turning it into a critical contention point and creating a performance bottleneck. We will see similar effects in our experiments later.

\section{Multi-Striding Analysis}
\label{sec:analysis}
\begin{figure*}
    \centering
    \includegraphics[width=0.95\linewidth]{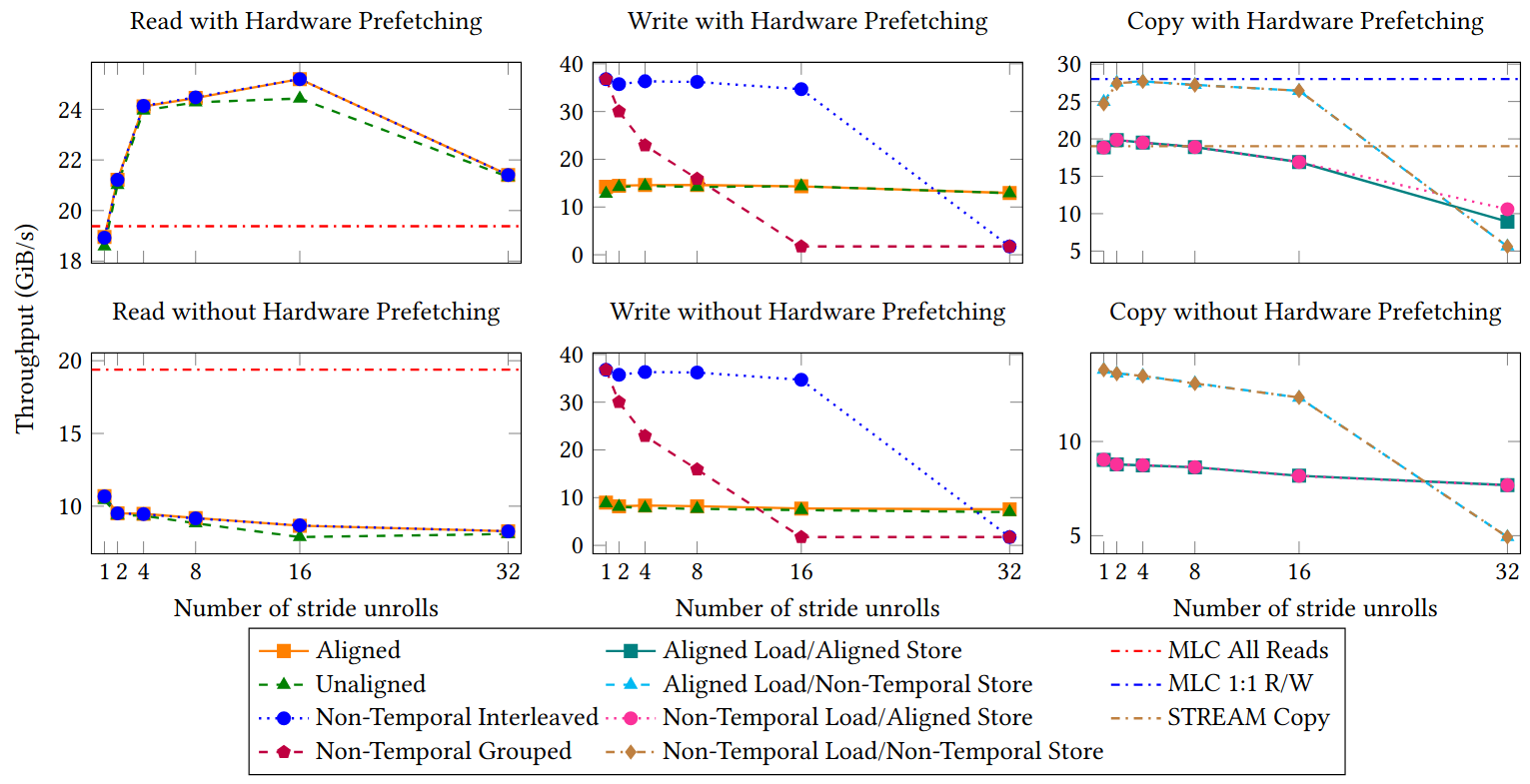}
\caption{Measured throughput of different memory operations for increasing numbers of strides on the Core i7-8700 platform.}
\label{fig:read_write_copy}
\end{figure*}


In this section, we characterize the performance of the memory subsystem when using multi-strided access patterns. 
We show that multi-striding results in improved single-core memory throughput thanks to better exploitation of the hardware prefetcher.
To this end, we have developed a set of specific micro-benchmarks.

\subsection{Experimental Design}
To establish the effectiveness of multi-striding we first measure the achieved memory throughput when reading a long sequence of consecutive data 
using various numbers of strides. We compare to the single-strided performance as a baseline. The micro-benchmark consists of a single loop that processes the data stored in an array using solely data-movement instructions.
In all cases, the loop uses an addressing mode that adds an immediate offset to the contents of a base 
register, which is incremented over loop iterations, similar to how many compilers generate such code.
Stride and portion unrolled code yield many possible arrangements of these accesses within the loop body.
The instructions in the loop bodies of micro-benchmarks are arranged in a grouped manner, as opposed to an interleaved manner, meaning that all accesses to the same stride are done consecutively, 
before proceeding to the next stride.
This arrangement of instructions showed higher throughput, though the final execution order is dependent on the CPU.

As described before, the number of stride unrolls that can be attained depends on the loop unroll factor. However, varying loop unroll factors result in loop bodies with different lengths, loop iteration counts and amounts of executed branch instructions. All of these have an influence on the achieved performance and memory bandwidth. 
Therefore, we directly generate AVX2 assembly code and 
enforce a constant number of 32 loop body unrolls. These 32 unroll slots are evenly distributed over the given number of stride unrolls in each configuration. By doing so, the only differences between configurations of the micro-benchmark are: (\emph{1}.) the offsets at which each instruction accesses data and (\emph{2}.) the step-size by which we increment the base register. 
This way, we completely isolate the effects of multi-striding to obtain an independent analysis.

We consider two array sizes to probe the effect of set placement. One array size is not an exact power of two, whereas the other is.
The size of the array must be well beyond the size of the system's L3 cache available to a single core. We chose sizes of approximately $1.9$ and exactly $2.0$ GiB, as these are larger than the available L3 cache of 12 MiB while remaining addressable using a base register combined with a signed 32-bit immediate.
We hypothesize that a regular distribution of strides over an array with a size of exactly a power of two causes the multiple strides accessed within a loop iteration to hit the same cache sets, resulting in many collisions.
In contrast, the non-power-of-two $1.9$ GiB array in the experiments should lead to a more evenly distribution of cache sets over the strides at any given iteration. We test our hypothesis by exploring the performance difference between the two array sizes in Section~\ref{sec:cache-collisions}.


We generated many configurations of this micro-benchmark that only differ in the following aspects:
\begin{enumerate}[wide,noitemsep,labelwidth=!,labelindent=0pt]
\item The instruction under test: load, store, or a combination thereof using different AVX2 instructions (see the preceding section).
\item The number of stride unrolls present in the access pattern generated by the loop.
\end{enumerate}
\noindent Because of our focus on regular vector load/store instructions to investigate the effect on maximum achieved memory bandwidth, we will not consider broadcast, scatter and gather instructions.


\subsection{Experimental Setup}
We conducted experiments 
on the Intel Coffee Lake 
micro-architecture, for which the specifications are given in Table~\ref{tab:specs}.
To accurately characterize the hardware prefetcher behavior, we carefully configured this machine to rule out as many other effects as possible: we locked the CPU frequency to $3.2$ GHz, turned off TurboBoost, HyperThreading, limited C-states (to disable deep sleep states) and enabled huge pages.
The CPU allows hardware prefetching to be enabled and disabled through its Model-Specific Register (MSR).
In our experiments hardware prefetching is enabled unless otherwise noted.

To measure the memory throughput, we first perform two warm-up executions, after which we measure the execution time over 10 executions, and take the median over 5 such runs. All loads and stores are enforced to be executed before we stop measuring the execution time by using a memory fence instruction after the loop.

We compare to a single-strided 32 loop unrolled baseline. Moreover, we included memory bandwidth measurements from established benchmarks as a reference. We consider the ``Copy'' benchmark from STREAM version 5.10~\cite{McCalpin1995, McCalpin2007} (modified to make use of huge pages as well) compiled with CLang version 20.0 and the ``ALL Reads'' and ``1:1 Reads-Writes'' benchmarks from the Intel Memory Latency Checker binary version 3.11 (MLC)~\cite{MLC}, both using huge pages and forced to execute on a single core.

\subsection{Read Performance}
We first explore read performance for an increasing number of stride unrolls while carefully validating that the effects are indeed due to improved hardware prefetcher utilization.
We ran micro-benchmarks that use different types of load instructions: aligned, (4-byte offset) unaligned and non-temporal instructions.
The achieved memory throughput is shown in the top left graph of Figure~\ref{fig:read_write_copy}.
We observe a clear increase in memory throughput for all surveyed types of load instructions when multiple strides are used.
More specifically, at 16 strides, aligned and non-temporal loads show an improvement of 33\% and unaligned loads of 31\% over the single-strided baseline.
Our single-strided baseline is about on par with the Intel MLC benchmark (the horizontal line in red), which confirms that our baseline achieves competitive performance. Importantly, our multi-strided benchmarks obtain higher throughput than MLC, affirming we obtain higher bandwidths thanks to multi-striding.

\begin{figure}
    \centering
    \includegraphics[width=0.95\linewidth]{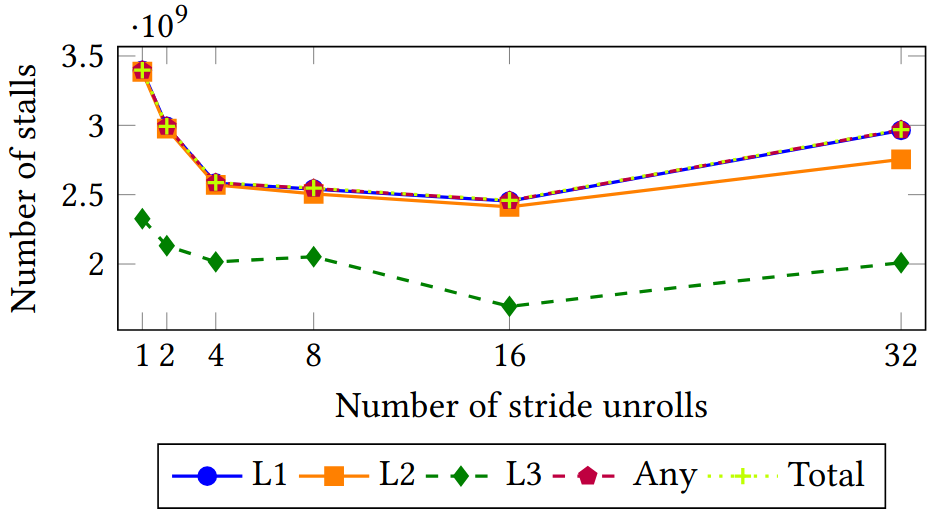}
\caption{Execution stalls with outstanding loads for L1 data, L2 and L3 cache misses, any outstanding load and the total number of stalls.}
\label{fig:stalls}
\end{figure}

We hypothesize that the increase in throughput realized by our multi-striding technique is due to the better exploitation of the hardware prefetcher.
To investigate, we first consider the number of memory stall cycles for different stride counts. Since the micro-benchmarks are fully memory-bound, we expect stall cycles to decrease as the number of stride unrolls increases.
Figure~\ref{fig:stalls} shows the number of memory stall cycles for outstanding loads at different levels of the cache hierarchy, measured using \texttt{perf}.
The total number of stalls follows a pattern similar to the throughput, suggesting a correlation between the two.
The lines for execution stalls related to L1 cache misses and any outstanding loads align closely with the total number of stalls.
This indicates that nearly all execution stalls coincided with outstanding memory loads, particularly originating from instances of L1 cache misses.
For stride configurations with less than 8 strides, nearly every stall cycle involves at least one L2 cache miss, before becoming relatively less frequent (ultimately 92\% at 32 strides) as we increase the number of stride unrolls, which can be seen from the growing distance to the total.
The distance between L3 and the total starts at 68\% for a single-strided configuration, then becomes \emph{relatively} more frequent at 8 strides with 81\% and declines to 68\% of the total again.
With cache misses at the L3 level becoming more frequent than those at L2, requested data can be found more commonly in L1 and/or L2 in these cases.



\begin{figure}
    \centering
    \includegraphics[width=0.95\linewidth]{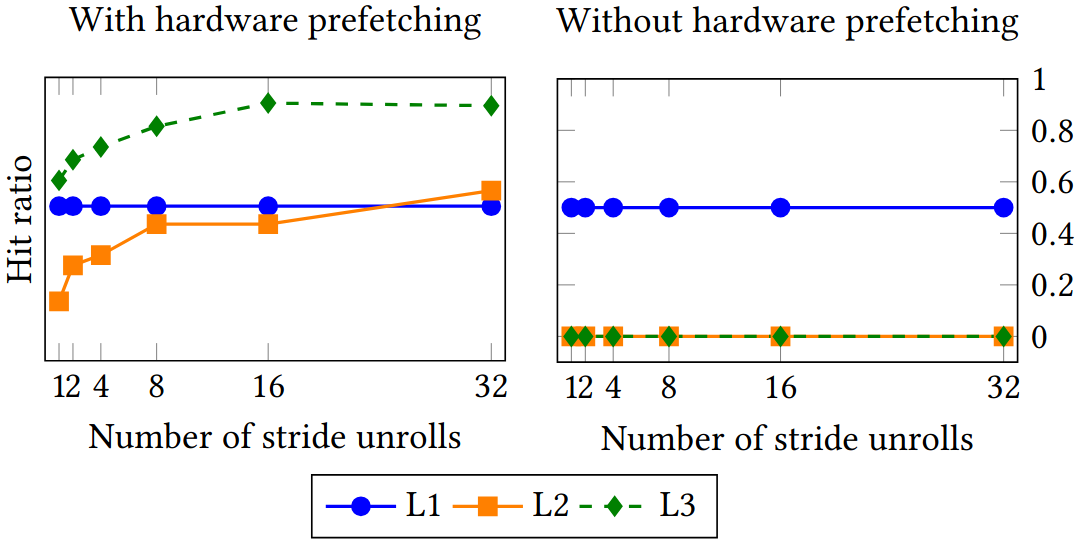}
    \caption{Cache hit ratio for different cache levels with and without hardware prefetching.}
\label{fig:read_hitratio}
\end{figure}

Together with the reduction of stall cycles we also expect an improvement in the cache hit ratios if this is due to better prefetcher performance.
Figure~\ref{fig:read_hitratio} shows the cache hit ratio for different cache levels using different numbers of strides, again obtained using \texttt{perf}.
On the left, with hardware prefetching enabled, the probability of finding the requested data in the L2 and L3 layers of the cache increases as the number of stride unrolls is increased. This correlates with the decrease in stall cycles.
Since there is no data reuse in our micro-benchmarks, this data cannot be present in the cache from an earlier access, indicating the data has been prefetched.
The hit ratio for L1 cache is exactly $0.5$, 
because the data is consumed faster than the prefetcher can put it in this cache layer, causing a miss on the first half of the block and experiencing a hit on every second half of -- the now included -- corresponding cache line.
The difference in L1 outstanding load stall cycles, despite its hit ratio being stable, is explained by its dependency on L2 and L3 hit ratios.
More specifically, the number of these stall cycles depends on
in which level of the cache hierarchy the data is found.
We conclude that the use of multiple strides primes the prefetcher to prefetch data more efficiently in the L2 and L3 layers, increasing cache utilization and memory bandwidth.

Finally, to verify that these improvements are due to better utilization of the hardware prefetcher, we test with the hardware prefetcher disabled.
And indeed, as can be seen in the bottom left graph of Figure~\ref{fig:read_write_copy}, no improvement occurs in this case.
Rather, the achieved memory bandwidth is even reduced if the number of stride unrolls is increased.
The graph on the right in Figure~\ref{fig:read_hitratio} also confirms this. The hit ratios for L2 and L3 are zero, signifying that the necessary data is never brought to cache ahead of time by a prefetcher. We 
conclude that the hardware prefetcher is responsible for placing the data in cache ahead of time and that these effects are improved when multiple strides are utilized.

\subsection{Write Performance}
Next, we consider the performance of the aforementioned write instructions. 
Additionally, we benchmark non-temporal writes that use an alternative arrangement of instructions in the loop body. In this case we visit each stride after one another before proceeding to the next offset, 
causing interleaved accesses over strides.

The results are depicted in the top middle graph in Figure~\ref{fig:read_write_copy}.
We observe that the write benchmark exposes the same effect, where multi-strided aligned stores see a improvement of 3\% and multi-strided unaligned stores 
of 13\% over the single-strided baselines.
Again, as seen in the bottom middle figure, there is no improvement with the hardware prefetcher disabled, but rather a slight decline.
Non-temporal writes follow a different pattern and also seem to be unaffected by the hardware prefetcher; multi-strided kernels do not surpass the throughput of the single-strided kernel. 
These operations 
seem only beneficial when sequences of 
it are arranged in groups of consecutive accesses, rather than interleaved among strides, presumably due to write buffer merging.
According to Intel documentation, the non-temporal write instructions in this micro-architecture are no-write-allocate. 
Thus, contrary to the other write instructions, it does not depend on the cache block being prefetched, as the block is not allocated to cache but written directly to memory.
We hypothesize the low performance when using many strides is caused by overwhelming the write-buffer with an increasing number of different simultaneously accessed cache blocks, turning it into a 
bottleneck in the path from the register to 
memory. 
This contention puts a 
limit of about 1.74 GiB/s on the overall throughput, as can be noticed from the equally poor performance of the 16 and 32 stride configurations for interleaved accesses.

\subsection{The Effect of Cache Collisions}
\label{sec:cache-collisions}
\begin{figure}
    \centering
    \includegraphics[width=0.95\linewidth]{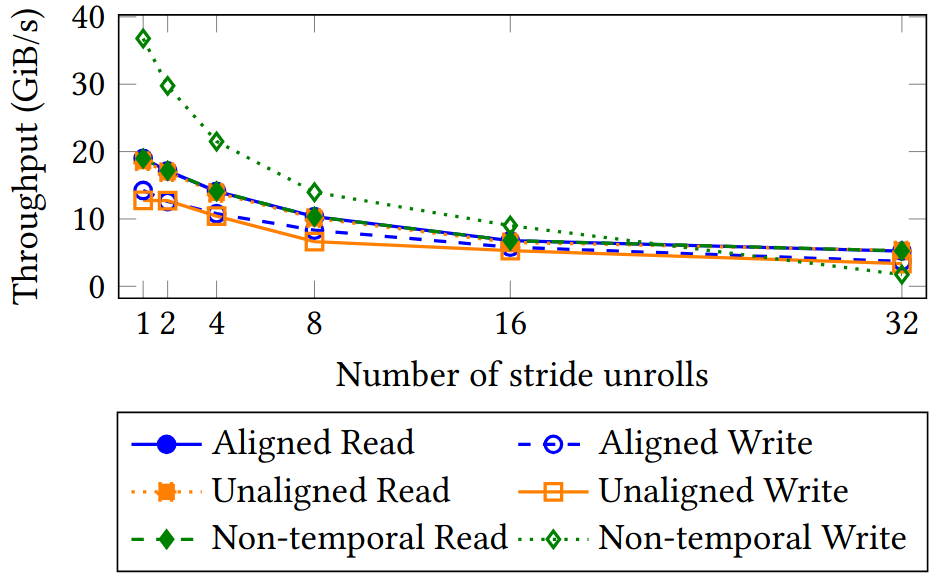}
\caption{Throughput in gigibytes per second using different types of data-movement instructions for different stride configurations on exactly 2 GiB of data.}
\label{fig:aligned}
\end{figure}

To assess how the distribution of strides over a certain number of cache sets impacts performance, we executed the micro-benchmarks with exactly $2$ GiB of data instead of $1.9$ GiB. Understanding this effect helps us in designing a generic methodology for transforming compute kernels to use multiple strides.
The results of this experiment for read and write instructions are shown in Figure~\ref{fig:aligned}.
The throughput of each single-strided kernel 
rapidly declines as the number of stride unrolls increases.
So, for multi-striding to be effective, the distribution of strides over the datastructure must be carefully chosen.
We expect the observed drop in throughput to be caused by the fact that instructions within the same loop iteration perform memory accesses that fall into the same cache set. As a result, the different strides will be competing for ways within the same set.

To further investigate this we consider cache statistics obtained using \texttt{perf}.
The number of L3 misses is doubled at 8 strides as opposed to the earlier experiment using $1.9$ GiB of data, up to a 560\% increase for kernels using 32 strides.
The L3 hit ratio for this setup is only 80\% at most for 32 strides, as opposed to the 90\% found from 16 strides in the earlier setup.
Furthermore, the number of stall cycles is doubled for the 4-strided kernel relative to the previous setup, and is increased by 477\% for 32 strides.
These stall cycles involve L3 cache misses in 90\% of the cases already when 4 strides are being used.
In conclusion, a distribution of simultaneous accesses corresponding to the same set is unfavorable as it results in a surge of cache misses, a lower hit ratio and, in turn, more execution stalls involving outstanding loads.
For effective multi-striding memory accesses within the same loop iteration must fall into different cache sets.



\subsection{Read-Write Performance}

Finally, we examine the combination of reads and writes within the loop body.
In the copy micro-benchmark, we explore different configurations of aligned and non-temporal loads and stores.
The results are shown in the top right graph of Figure~\ref{fig:read_write_copy}.
We notice that multi-strided kernels up to 16 strides achieve a performance increase of 11\% over single-strided kernels when using non-temporal stores.
For aligned stores this is the case up to 8 strides with an increase of 5\%.
A decline arises with fewer strides than seen before, as in reality, each stride in these kernels results in an access pattern with two access sequences: one for the read data and one for the written data, effectively doubling the number of patterns to be recognized by the prefetcher.

In the same figure can be seen that our single-strided copy benchmarks achieve results comparable to the MLC and STREAM benchmarks, whereas our multi-strided benchmarks show increased performance.
Again, note that no performance improvements are observed when disabling the hardware prefetcher (bottom right), instead the performance declines in this case as well. This stipulates that the achieved improvements are indeed due to better exploitation of the hardware prefetcher.

\section{Compute kernel transformation}
\label{sec:transformation}
We now describe how we applied multi-striding on several compute kernels.
The overall methodology of the transformation is described first, followed by a concrete example.

\subsection{Overall Methodology}
Within this paper we consider compute kernels with regular access patterns on dense data. 
The computational loops in these kernels must be free of (loop-carried) data dependencies that enforce a fixed order of execution or limit the use of vectorized updates.
This allows sequences of data to be freely distributed over vectorized strides, which is a requirement for successful application of multi-striding.

\subsubsection{Preparatory Transformation}
The first step is to determine in which direction vectorization and multi-striding are to be applied. We refer to this direction as the \emph{contiguous data axis}.
We presume (without loss of generality) that arrays are stored in row-major order.
As multi-striding optimizes for memory bandwidth, the aim is to select the array access that will be most demanding in terms of bandwidth, under the assumption that this memory access bottlenecks the execution.
We will refer to this as the \emph{critical memory access}.
The critical memory access is found by selecting the datastructure with the highest dimensionality, for which holds that the last indexing variable used in this access appears exclusively as the last dimension in every array indexed with that variable.
For now, we view the array with the highest dimensionality as the one that will accrue the least amount of reuse.
We select the last dimension of the found array as the contiguous data axis. This implies that we will be vectorizing the loop such that consecutive loop iterations access adjacent memory locations.

The condition on the indexing variable is crucial as it disallows vectors to be gathered from several memory locations.
Consider a matrix transpose kernel (\verb|A[i][j] = B[j][i]|). If we select \texttt{B} as main array, array accesses \texttt{B[j][i]} are vectorized over \texttt{i} and made adjacent.
However, in the access to \texttt{A}, the index \texttt{i} is not in the last position, implying that vectorizing over \texttt{i} requires values from \texttt{A} to be gathered from different rows. This results in multiple additional strides and requires bringing more blocks to the cache. This makes it less reliable for the prefetcher to detect the indented strides.
As mentioned before, access patterns involving gather instructions are not considered further within this work.

Subsequently, the loop is vectorized along the contiguous data axis. For example, if \texttt{B[j][i]} is selected as the critical memory access, the aim is to perform the operations on vectors \texttt{B[j][i:i+w]}, where \texttt{w} is the vector width in the number of elements. First, we ensure that the innermost loop iterates over the chosen axis (\texttt{i} in case of \texttt{B[j][i]}).
If this is not the case, we first apply loop interchange, which can be done in all cases since loops must be free of loop-carried dependencies.
Secondly, we vectorize the loop.
In case a kernel traverses a one-dimensional array, this array must first be distributed into $n$ partitions when $n$ strides are to be generated. The loop blocking transformation is used to achieve this.


\subsubsection{Code Generation}
To ensure full control over the generated instructions, we directly generate (AVX2) assembly for the transformed loop by means of a series of Python scripts. We use addressing modes that add an immediate offset to the contents of a base register $B$ that is incremented over loop iterations. 


The resulting loop has a large optimization space. To be able to explore this optimization space, we write a parametrized assembly template and use a series of Python scripts to instantiate many configurations of a compute kernel for different parameters. A configuration is defined by the \emph{portion unroll} and \emph{stride unroll} parameters.
\emph{Portion unroll} controls the portion of each stride that is processed within each loop iteration; \emph{stride unroll} defines the number of (concurrent) strides by unrolling the \emph{outer loops} to increase this number.
From such a configuration, the parameters for the assembly template can be computed: the offsets of the accesses, step sizes and the total number of unrolls.
Both unrolling methods adjust the step size in their respective dimensions, which are selected as the largest numbers divisible by those step sizes within set limits. This prevents the need to process leftover array parts separately or exceed bounds.


Portion and stride unrolling may result in redundant loads and stores to be emitted. 
We implemented additional Python scripts to generate assembly code with these redundant instructions eliminated as an optimization.
This introduces restrictions regarding the number of unrolls, because due to the elimination of intermediate load and store instructions, more values have to remain in registers throughout the entire loop execution.
Striding configurations that require more registers than are available in a particular architecture are considered infeasible and are excluded.
We avoid register spilling, as this would only reintroduce some of the removed redundant loads and stores.

\subsection{Example}
As an example of the application of multi-striding, consider the simple transposed matrix vector multiplication kernel in Listing~\ref{lst:before} which is similar to the gemvermxv1 and doitgen kernels that will be examined in the next section.
We identify \texttt{A[j][i]} to be the critical memory access, as from the dimensionality of \texttt{A} can be deduced that it will contain many more values than \texttt{B} and \texttt{C}, and its values are not reused within this loop.
The variable \texttt{i} is used to index the last dimension of \texttt{A} and is never used as index to a non-last dimension. Therefore, we will transform the loop to vectorize along \texttt{i} as contiguous data axis. In this case, we must perform loop interchange  first to ensure subsequent iterations of the innermost loop access adjacent values of \texttt{A[j][i]} in memory. After that, we vectorize the loop and create a parametrized assembly template, in our case for groups of 8 single-precision floats using AVX2.

\begin{figure}[b]
\setcaptiontype{lstlisting}%
\lstset{
    basicstyle=\footnotesize,
    aboveskip=-5pt,
    belowskip=-5pt
}
\begin{lstlisting}[language=C]
for (int i = 0; i < N; i++) {
    for (int j = 0; j < M; j++) {
        C[i] += A[j][i] * B[j];
    }
}
\end{lstlisting}%
\caption{Transposed matrix-vector multiplication.}
\label{lst:before}
\end{figure} 
\begin{figure}[b]
\setcaptiontype{lstlisting}%
\lstset{
    basicstyle=\footnotesize,
    aboveskip=-10pt,
    belowskip=-10pt
}\begin{lstlisting}[language=C]
for (int j = 0; j < M; j += 3) {
    for (int i = 0; i < N; i += 8*2) {
        C[i:i+8] += A[j][i:i+8] * B[j];
        C[i:i+8] += A[j+1][i:i+8] * B[j+1];
        C[i:i+8] += A[j+2][i:i+8] * B[j+2];
        C[i+8:i+16] += A[j][i+8:i+16] * B[j];
        C[i+8:i+16] += A[j+ 1][i+8:i+16] * B[j+1];
        C[i+8:i+16] += A[j+ 2][i+8:i+16] * B[j+2];
    }
}
\end{lstlisting}
\caption{Multi-strided matrix-vector multiplication.}
\label{lst:after}
\end{figure}

Configurations of the assembly template can be instantiated by specifying unroll factors for portion unroll and stride unroll. Listing~\ref{lst:after} presents a C-like representation of the instantiated assembly template for a portion unroll factor of $2$ (over \texttt{i}) and a stride unroll factor of $3$ (over \texttt{j}), resulting in six unrolls distributed over two strides. Note that this configuration is only valid if \texttt{M} and \texttt{N} are multiples of their corresponding step sizes.


\section{Experiments}
\label{sec:experiments}
In this section, we first characterize the performance of multi-striding in practice on six compute kernels of which many configurations were generated. Secondly, we compare the performance of the best 
generated kernel to state-of-the-art implementations.
We conducted these experiments on three different micro-architectures to establish that multi-striding can be commonly applied.

\subsection{Surveyed Kernels}
\begin{table*}[h!]
\centering
\caption{Overview of the six surveyed compute kernels used in our experimentation.}
\label{tab:sizes}
\footnotesize
\begin{tabular}{|l|l||c||l|l|l||c|c|c|c||l|c|c|}\hline
 \multicolumn{2}{|c|}{\textbf{Kernel}}& \textbf{AT}& \multicolumn{3}{|c|}{\textbf{Strides}} & \multicolumn{5}{|c|}{\textbf{Characteristics}}& \multicolumn{2}{|c|}{\textbf{Data size (GiB)}}\\\hline
\textbf{Name}&\textbf{Description}&&\textbf{L}& \textbf{S}&\textbf{L/S} & \textbf{IN}&\textbf{WB} &\textbf{LE} &\textbf{LI} &\textbf{LB}& \textbf{Isolated}& \textbf{Comparison}\\\hline \hline
bicg*&BiCG Sub Kernel of BiCGStab Linear Solver &A& $n+2$& $1$&$1$ & Y& & & && 4                                      & 4                                   \\ \hline
conv                   &3x3 2D Convolution Stencil &U& $n+2$& $n$& & & & & && 2& 2\\ \hline
doitgen*&Multi-resolution analysis kernel (MADNESS) &A& $n+1$& &1 & Y& Y&2 &Y && 4                                      & 0.4\\ \hline
gemver*&Vector Multiplication and Matrix Addition &A& & & & & & & && & 4                                   \\ \hline 
- gemverouter*&  Double Rank-1 Matrix Update& A& $4$& & $n$ & & & & && 4&4\\ \hline 
- gemvermxv1*&  Transpose Matrix Vector Multiplication& A& $n+1$& & $1$ & & & &Y && 4&4\\ \hline 
- gemversum*& Vector Sum Update& A& $n$& $n$&  & & & & &Y& 4&4\\ \hline 
- gemvermxv2*& Matrix Vector Multiplication& A& $n+1$& & $1$ & & & & && 4&4\\\hline
jacobi2d*&2D Jacobi Stencil &U& $n+2$& $n$& & & Y&1 & && 2& 2                                   \\ \hline
mxv&Matrix Vector Multiplication &A& $n+1$& &1 & & & & && 4                                      & 4                                   \\ \hline
init&Initialization    &A& & $n$& & & & & &Y& 2                                      & 2                                   \\ \hline
writeback&Writeback    &A& $n$& $n$& & & & & &Y& 2                                      & 2                                   \\ \hline
\end{tabular}
\end{table*}
We explore six memory-bound kernels that are outlined in Table~\ref{tab:sizes}. The kernels marked with an asterisk are taken from PolyBench~\cite{polybench}. The gemver kernel consists of four separate steps that will be considered individually. Additionally, \emph{init} and \emph{writeback} kernels are included because some kernels include an initialization or write back phase as shown in the table.
The access type column (AT) shows whether the generated kernel uses aligned (A) or unaligned (U) memory access. 
Unaligned access is used for both stencil computations, as they involve padding, which offsets strides by values that break 32-byte alignment required for aligned instructions.
The strides column describes the memory access pattern generated by a kernel in terms of the number of load (L), store (S) and load/store (L/S) strides, where $n$ is the number of \emph{stride unrolls}. The table shows that we indeed selected  kernels that generate diverse memory access patterns.
Loop interchange (LI) and loop blocking (LB) indicate whether these transformations were applied during the transformation of the compute kernel to a multi-strided variant.


These kernels will be used in the experiments presented in Sections~\ref{sec:striding_in_practice} and~\ref{sec:comparison}.
In the first set of experiments, we are interested in learning how effective the application of multi-striding is for the essential computation performed by each compute kernel.
However, compute kernels typically consist of multiple separate steps which exhibit different memory access patterns. 
To test the effect of a particular multi-striding configuration without interference of the other steps, as indicated by the initialization (IN), writeback (WB) and Loop Embedment (LE) columns, we split these separate steps into isolated kernels.
More specifically, for doitgen and jacobi2d the unnecessary outer loops were removed as well as the initialization and writeback steps in general.
To ensure comparable loop bodies in terms of amount and type of instructions regardless of the striding configuration, the loads and stores from each unroll are performed, even when redundant.
As we empirically assessed aligned and interleaved accesses showed higher throughput than non-temporal and grouped accesses in all kernels, we use the former in these loop bodies for both experiments.
For the second set of experiments, where we compare to the state-of-the-art, we reassemble the compute kernels using the best striding configuration found for each of the separate steps.
We use 32-bit immediate addressing and problem sizes between 2 - 4 GiB such that these are divisible by the respective step sizes and facilitate aligned data movement (except for conv and jacobi2d which perform unaligned operations).

\subsection{Experimental Setup}
\begin{table}
    \centering
    \caption{Specifications of used micro-architectures. }
    \label{tab:specs}
    \footnotesize
    \begin{tabular}{|l||c|c|c|}\hline
 & \multicolumn{3}{|c|}{\textbf{Micro-architecture}}\\\hline
                           & \textbf{Coffee Lake}   & \textbf{Cascade Lake}  & \textbf{Zen 2} \\ \hline\hline
        Vendor             & Intel         & Intel      & AMD \\\hline
        Model family       & Core          & Xeon Silver  & EPYC\\\hline
        Model name         & i7-8700        & 4214R  & 7402P \\\hline
        Operating System   & Rocky Linux  8.10 & Rocky Linux 8.5 & CentOS Linux 7\\\hline
        Base/Max/Min Freq. & 3.2/4.6/0.8 GHz & 2.4/3.5/1.0 GHz & 2.8/3.2/1.2 GHz \\ \hline
        Bandwidth (GiB/s)& 19.87 & 17.88 & 23.84 \\ \hline 
 Memory channels& 2 & 6 & 8\\\hline
        L1 Inst. size/assoc. & 32 KiB / 8-way & 32 KiB / 8-way & 32 KiB / 8-way \\\hline
        L1 Data size/assoc.  & 32 KiB / 8-way & 32 KiB / 8-way & 32 KiB / 8-way \\\hline
        L2 size/assoc.      & 256 KiB / 4-way & 1 MiB / 16-way & 512 KiB / 8-way \\\hline
        L3 size/assoc.      & 12 MiB / 16-way  & 16.5 MiB / 11-way & 16 MiB / 16-way \\\hline
        RAM size           & 16 GiB         & 256 GiB     & 128 GiB\\\hline
        Max. FMA   & 147.2 GFLOP/s & 112.0 GFLOP/s & 102.4 GFLOP/s \\\hline
    \end{tabular}
\end{table}





We consider three different micro-architectures that are outlined in Table~\ref{tab:specs}, each using their default page size (4 KiB).
The L3 is shared among all cores, except for Zen 2.
All caches have a cache line size of 64 bytes.
On all machines, we use LLVM, CLang and Polly version 20.0.0, Halide version 18.0.0, MKL version 2024.2, OpenBLAS version 0.3.28 and OpenCV version 4.10.0.
The implementation of our kernels use the AVX2 instruction set extension using single-precision floating points (8 values per register).
For each experiment, we collect the median over 5 measurements, each including 5 executions.
At the very end of each compute kernel, we place an \texttt{mfence} instruction to ensure all loads and stores are performed before measurement is stopped.

\subsection{Characterization of Multi-Striding}
\label{sec:striding_in_practice}
To characterize the effectiveness of multi-striding, we explore the optimization space of multi-striding for the isolated computational steps of the described compute kernels.
We evenly distribute a given total number of unrolls (1 up to 50) over a number of stride unrolls and portion unrolls.
This results in a large number of configurations and for each configuration we generate an AVX2-executable using our methodology described in Section~\ref{sec:approach}.
Note that for each number of stride unrolls we have multiple data points, namely one for each portion unrolled configuration.
For example, with 6 unrolls, we can have 1 stride of 6 vectors processed each iteration, 2 strides of length 3, 3 strides of length 2, or 6 strides of length 1.
Because each of the stride configurations will have different divisibility constraints on the dimensions of the data, we report throughput rather than time to compare kernels operating on data of different sizes.

\begin{figure*}
    \centering \includegraphics[width=0.86\linewidth]{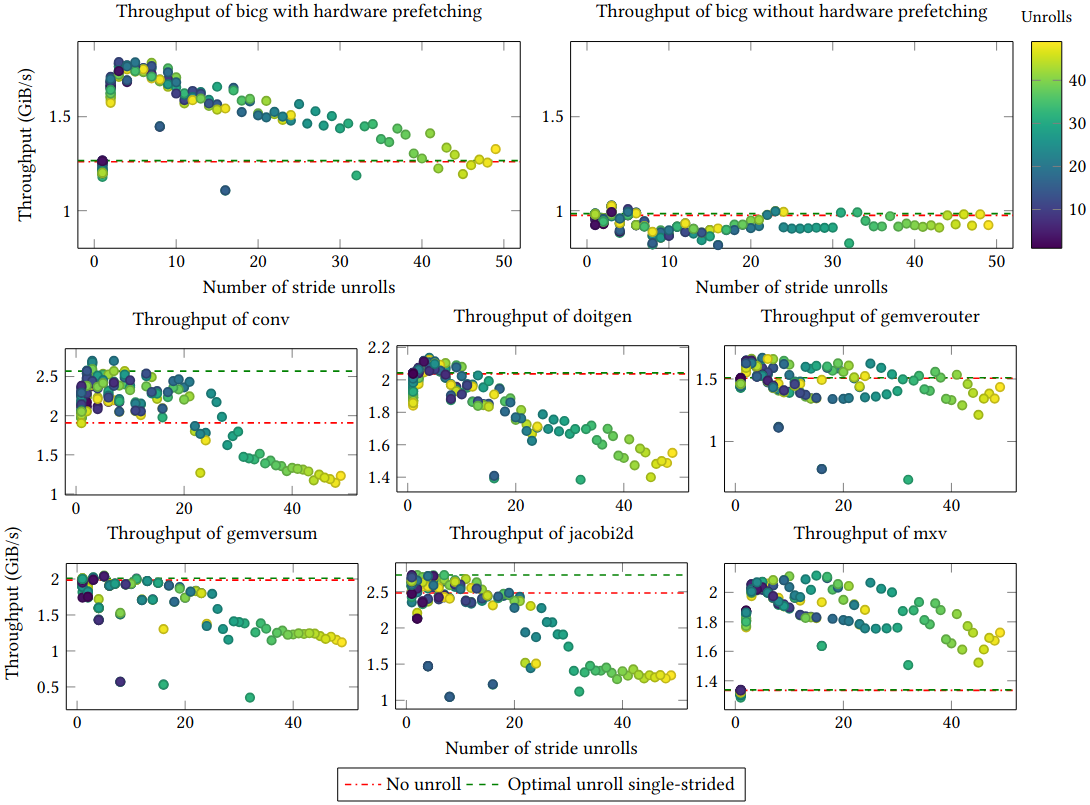}
\caption{Throughput of different isolated parts of compute kernels for different striding configurations.}
\label{fig:kernels}
\end{figure*}



In Figure~\ref{fig:kernels} we plot the throughput of the different (isolated) compute kernels as a function of the number of stride unrolls.
The green horizontal line shows our baseline: the best performing single-strided kernel that uses only loop unrolls in the contiguous data axis.
The horizontal line in red indicates the throughput of the kernel that uses no unrolls.
The data points in the figures are colored with a gradient, according to the \emph{total} number of unrolls.
For brevity, we have left out the figures corresponding to the two matrix-vector multiplication operations of gemver, as they are equivalent to the mxv and loop interchanged doitgen kernels.

The upper left and upper right graph show the throughput for the configurations for the bicg kernel with and without hardware prefetching, respectively.
From these results, it immediately becomes apparent that increasing the number of strides has a positive effect when hardware prefetching is enabled and has no significant effect otherwise. This reaffirms that multi-striding takes better advantage of the available hardware prefetchers.

Hardware prefetching was enabled in all other results. There is a consistent trend among the different compute kernels, where a number of stride unrolls between 1 and 10 provides the highest throughput, before declining as this number increases.
The number of total strides does not have a significant effect on the throughput, as indicated by the gradient.
Striding configurations at multiples of eight show outliers, explained by cache collisions (Section~\ref{sec:cache-collisions}).

All results positioned above the green line surpass the best baseline configuration, showing multi-striding has a beneficial effect over default loop unrolling in the contiguous data axis.
As shown by the red line, ignoring loop unrolling leaves additional performance gains untapped.
The jacobi2d kernel is an exception: the best multi-strided variants perform equally well as, but not better than, the best single-strided configuration.
The conv kernel, with an access pattern similar to jacobi2d, uses fused multiply-add instructions, while jacobi2d relies mainly on additions, suggesting a dependency on computation type.
Aside from this exception, these results demonstrate how properly chosen kernel variations using our proposed multi-striding technique significantly outperform the single-strided baselines, with observed speedups from a factor 1.02x for gemversum to a factor 1.58x for mxv.

\subsection{Comparison with the State-of-the-Art}
\label{sec:comparison}
\begin{figure*}
    \centering
    \includegraphics[width=0.9\linewidth]{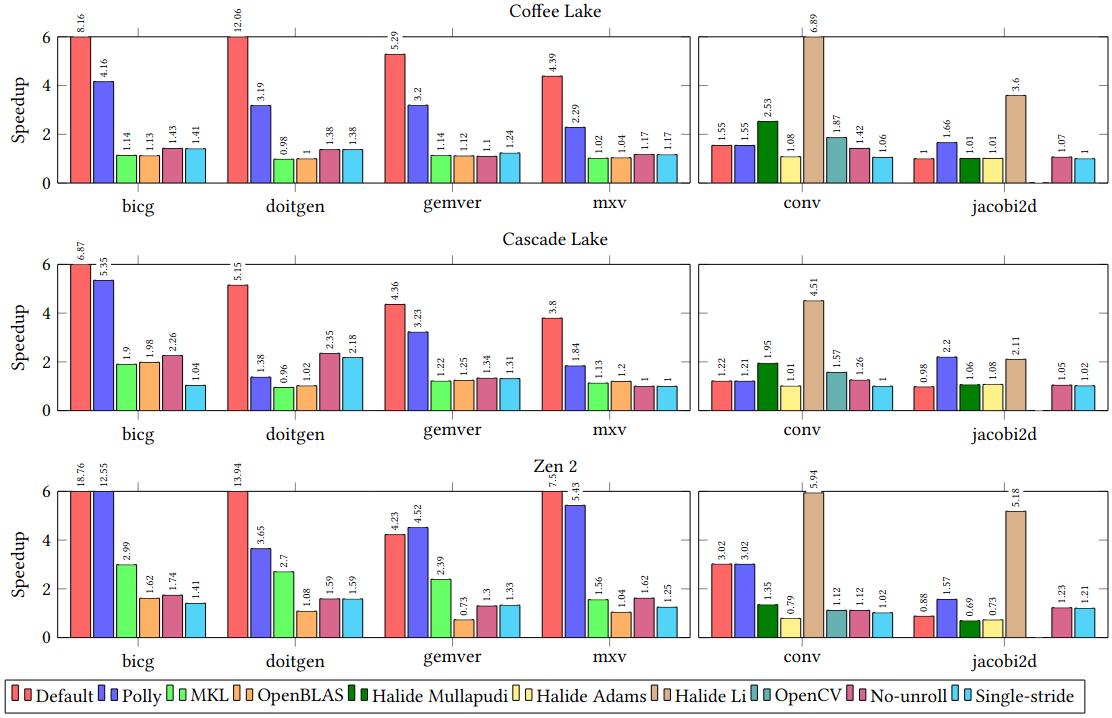}
\caption{Speedup in throughput of the best multi-strided configuration of the six compute kernels compared to state-of-the-art reference implementations, and single-strided and non-unrolled implementations on three micro-architectures.}
\label{fig:comparison}
\end{figure*}



Now that we have confirmed that multi-striding is able to leverage the hardware prefetcher to obtain significant speedups over single-strided configurations for various compute kernels, we will show that our multi-strided compute kernels in many cases outperform state-of-the-art reference kernels.



To reduce the parameter space of kernels consisting of multiple parts, we optimize each part individually in their performance exploration.
For example, we configure each the four individual parts of the gemver kernel using their highest-performing configurations found in isolated tests and unify these into a single configuration. 

Similarly, we apply the best configurations for initialization and writeback.
We have removed the redundant load and store operations in our generated compute kernels, but no elaborate optimizations have been performed.
For instance, in the jacobi2d kernel, some intermediate results are recomputed. Optimizations to counter this are unrelated to multi-striding and are therefore not considered.

For fair comparison, we display the highest recorded throughput among the used sizes of the data, per benchmark and micro-architecture.
We compare each of our top-performing multi-strided assembly implementations against a default CLang-optimized implementation, a CLang-optimized version with Polly and its stripmine vectorizer enabled, a generated assembly without unrolling, and the best-performing single-strided assembly implementation.
Furthermore, the bicg, doitgen, gemver and mxv kernels we compare with Intel's MKL and OpenBLAS.
The conv and jacobi2d stencil computations are compared with Halide, using the three auto-schedulers, Mullapudi~\cite{Mullapudi2016}, Adams~\cite{Adams2019} and Li~\cite{Li2018}.
The throughput of the conv kernel is also compared to OpenCV.
The use of AVX2 is verified, consistently accross machines only the bicg and mxv kernels for Polly and the mxv kernel made no use of it.
If a state-of-the-art implementation outperforms the best single-strided counterpart but is then surpassed by our best multi-strided implementation, it directly shows the value of multi-striding.

 

In Figure~\ref{fig:comparison} we find the speedup of multi-striding over different state-of-the-art implementations. 
Overall, our results show that compute kernels generated using our multi-striding methodology achieve noteworthy speedups over state-of-the-art hand-optimized kernels (e.g., Intel's MKL and OpenBLAS), for different micro-architectures, while only being slightly surpassed in a few cases. 
In many cases, state-of-the-art kernels outperform our best \emph{single-strided} implementation, while being surpassed by our \emph{multi-strided} variant. This shows that it indeed is multi-striding (and not loop-unrolling or hand written assembly) that increases performance beyond the state-of-the-art implementations. For jacobi2d, CLang outperforms multi-striding because it versions use shuffle and permutation instructions which we did not consider.
We conclude that multi-striding provides obvious advantages over state-of-the-art implementations and also can be generically applied.

\section{Conclusion}
\label{sec:conclusion}
In this paper, we have provided a thorough examination of multi-striding: a combination of loop transformations aimed to increase concurrent memory accesses in memory-bound kernels with regular access patterns. 
With our in-depth analysis, we showed that multi-striding achieves up to 2.18x higher memory throughput compared to single-strided baselines due to higher cache hit ratios as a result of better hardware prefetcher utilization.
%
%
We presented a comparison to the state-of-the-art using six memory-bound compute kernels. Our multi-strided kernels show significantly improved performance over code generated by contemporary compilers achieving speedups of up to 18.76x over CLang with vectorization enabled and 12.25x over CLang combined with Polly and its auto-vectorizer.
Additionally, we have shown that multi-striding also outperforms hand-optimized reference kernels: 2.99x over Intel's MKL, 1.98x over OpenBLAS, 1.08x over Halide (given a suitable autoscheduler is selected by the user) and 1.87x over OpenCV. 

Contrary to these reference kernels, the code for our multi-strided implementations is the result of a straightforward application of loop transformations and translation to SIMD-instructions, without any elaborate hand-optimization.
The methodology presented in this paper is therefore amendable to be implemented in compiler pipelines in the near future.

Given the impressive results, we aim to explore the application of multi-striding in other areas in future work.
As this paper focused on single-core performance, we will investigate how multi-striding performs in multi-threaded or multi-core settings, considering the shared resources in the memory hierarchy (e.g., L3 cache).
Furthermore, we plan to examine problems with irregular access patterns and those requiring gather/scatter instructions.
Finally, we foresee to implement our methodology in a compiler backend, such as integrating it in the LLVM compiler infrastructure.


\bibliographystyle{ACM-Reference-Format}
\bibliography{sample-base}

\end{document}